\documentclass[%
 reprint,
 amsmath,amssymb,
 aps,
]{revtex4-2}

\usepackage{graphicx}
\usepackage{dcolumn}
\usepackage{bm}

\usepackage{xcolor}
\usepackage{soul}
\usepackage{float}
\usepackage{subcaption}

\newcommand{\f}{\frac}

\newcommand{\p}{\partial}
\newcommand{\de}{\delta}
\newcommand{\g}{\gamma}

\newcommand{\om}{\omega}

\newcommand{\De}{\Delta}

\begin{document}

\preprint{APS/123-QED}

\title{Gaussian fluctuations of non-reciprocal systems}

\author{Sergei Shmakov}
 \email{sshmakov@uchicago.edu}
\affiliation{James Franck Institute and Department of Physics, The University of Chicago, Chicago, Illinois 60637, United States}
\author{Glasha Osipycheva}
 \email{osipychevag@uchicago.edu}
\affiliation{James Franck Institute and Department of Physics, The University of Chicago, Chicago, Illinois 60637, United States}
\author{Peter B. Littlewood}
 \email{littlewood@uchicago.edu}
\affiliation{James Franck Institute and Department of Physics, The University of Chicago, Chicago, Illinois 60637, United States}
\affiliation{School of Physics and Astronomy, The University of St Andrews, St Andrews, KY16 9AJ, United Kingdom}

\date{\today}

\begin{abstract}
Non-reciprocal systems can be thought of as disobeying Newton’s third law – an action does not cause an equal and opposite reaction. In recent years there has been a dramatic rise in interest towards such systems. On a fundamental level, they can be a basis of describing non-equilibrium and active states of matter, with applications ranging from physics to social sciences. However, often the first step to understanding complex nonlinear models is to linearize about the steady states. It is thus useful to develop a careful understanding of linear non-reciprocal systems, similar to our understanding of Gaussian systems in equilibrium statistical mechanics. In this work we explore simplest linear non-reciprocal models with noise and spatial extent. We describe their regions of stability and show how non-reciprocity can enhance the stability of a system. We demonstrate the appearance of exceptional and critical exceptional points with the respective enhancement of fluctuations for the latter. We show how strong non-reciprocity can lead to a finite-momentum instability. Finally, we comment how non-reciprocity can be a source of colored, 1/f type noise.   
\end{abstract}

\maketitle

\section{\label{sec:intro}Introduction}

Non-reciprocal interactions have received increased interest in recent years due to their wide applicability and theoretical insights for open systems and active matter. They arise in the context of dynamical phase transitions \cite{Fruchart2021, Anvi2024}, spin models \cite{Anvi2023,Anvi2024, Hanai2024}, models of neural dynamics with excitatory and inhibitory interactions \cite{Cook2022,Wilson1972,Dayan2001, calvo2024, Tian2022}, non-equilibrium quantum condensates \cite{Hanai2019}, among many others (for more examples see references within \cite{Fruchart2021}). Non-reciprocity pushes systems out of equilibrium, therefore producing novel phases of matter \cite{Fruchart2021}, critical exceptional points \cite{Zelle2023, Hanai2020}, and novel dynamics \cite{Weis2023, Shmakov2023}.  It is therefore useful to develop understanding of as many exactly solvable non-reciprocal models as possible, given especially the severe additional complexity that non-reciprocity can bring to an already complicated field of noisy nonlinear models.

In equilibrium statistical field theory, one of the few solvable systems is a Gaussian ensemble. From the perspective of dynamics, such an ensemble corresponds to a linear set of equations that define the relaxation of the order parameter to a stable steady state \cite{Hohenberg1977,Tauber2014, Altland2023}. Following that logic, we assume that a simple non-reciprocal model one ought to study is a set of linear stochastic equations. Statistical understanding of such a system can then be inferred either from promoting these equations to a path integral description through the MSR procedure \cite{Altland2023, Tauber2014, Kamenev2023, Martin1973, DeDominicis1978} or through the direct calculation of correlation functions. The linearized theory is often the starting point for stability analysis of a full nonlinear theory, for example to determine the upper critical dimension where the system can be described as mean field with small fluctuation corrections. 

We show below that increasing non-reciprocity expands the region of stability of the system with respect to the tuning of each field's relaxation constant. This is perhaps unsurprising: if the fields are coupled non-reciprocally, then an increase in one produces a decrease in the other, while the decrease in the other also produces the decrease in the first. That expansion of the stability region leads to an occurrence of exceptional lines (ELs) and critical exceptional points (CEPs). In dynamical systems exceptional points are marked by the merging of eigenvalues and eigenvectors (or Lyapunov vectors and exponents) of the matrix generating the dynamics \cite{Weis2023, Zelle2023}. In our case, exceptional lines will mark the onset of oscillations between the fields. CEPs will occur when the exceptional lines coincide with the critical lines, or when the point of two modes merging also happens to be critical. At these isolated points, the system will experience enhanced fluctuations. For the equilibrium Ising universality class, the critical point is marked by a $k^{-2}$ divergence in the fluctuation spectrum \cite{Altland2023,Kamenev2023,Tauber2014}. We observe a $k^{-4}$ divergence often associated with CEPs \cite{Hanai2020,Zelle2023}, and also get a $k^{-6}$ divergence happening at isolated points where several parameters need to be tuned.

Pattern formation or finite momentum instabilities are also prevalent in non-equilibrium systems \cite{Cross1993}. In reaction-diffusion systems, for instance, different temporal and diffusive constants are responsible for pattern-forming behavior. The overdamped models that we consider are a generalization of such linear models. Here we demonstrate the simple mechanism of forming finite momentum instabilities in the regime of strong non-reciprocity.

Finally, we show how non-reciprocal interactions lead to the appearance of $1/f$ noise. Noise and fluctuations with $1/f$ spectra are observed in a variety of systems with various explanations as to their origin \cite{Szendro2001, Dutta1981, Bak1987, milotti20021}. Here we show that such spectrum of noise can be achieved for the effective dynamics of one of the fields as the other is integrated out. The $1/f$ noise is generally observed for high enough frequencies, but can be extended to appear for all as the CEP is approached. 

This paper is organized as follows. In Sec. \ref{sec:overdamped} we introduce the simplest linear non-reciprocal model. We analyze its stability profile with respect to a measure of non-reciprocity that we define. We then proceed to write down the statistical theory, demonstrate the equilibrium nature of the critical lines as well as the enhanced fluctuations at the CEPs. We show how the finite momentum instability can easily occur in the regime of strong non-reciprocity and how the effective dynamics of a single field picks up $1/f$ noise due to the coupling to the other field. In Sec. \ref{sec:inertial}, we undergo the same steps, except for a model with inertial terms for the dynamics of the fields. We close with a discussion of results in Sec. \ref{sec:conclusion}.

\section{\label{sec:overdamped}Overdamped systems}
We start with a simple linear, spatially extended and noisy dynamical system:
\begin{equation}
    \begin{split}
        &\p_t \phi_1 = -m_1 \phi_1 + j_{12} \phi_2 + D_{1}\nabla^2 \phi_1 + \xi_1 \\
        &\p_t \phi_2 = -m_2 \phi_2 + j_{21} \phi_1 + D_{2}\nabla^2 \phi_2 + \xi_2
    \end{split} \; .
\end{equation}

Here (for $i \in {1,2}$) $\phi_i$ are scalar fields, $m_i$ are the relaxation constants of each field, $j_{12} \neq j_{21}$ are non-reciprocal couplings between the fields, $D_i$ are diffusion constants, and $\xi_i$ are Gaussian white noise variables with correlations $\left<\xi_i(x,t)\xi_j(x',t')\right> = B_{i/j}\de_{ij}\de(t-t')\de^d(x-x')$. We assumed that the noises are uncorrelated and that there is no cross-diffusion for simplicity. We pick the Ito discretization \cite{Kamenev2023,Tauber2014}.

This system is $\mathcal{Z}_2$ symmetric under the transformation $(\phi_1, \phi_2) \rightarrow (-\phi_1, -\phi_2)$. We therefore expect that an appropriate system for comparison is the $\mathcal{Z}_2$ transition in the Ising model from the disordered phase. Clearly, if we remove the cross-couplings $j_{12}$ and $j_{21}$, we get two copies of the equilibrium $\mathcal{Z}_2$ transition in the Gaussian approximation. In other words, this is two non-reciprocally coupled Model A systems \cite{Hohenberg1977, Tauber2014} in the linear approximation. The critical point in the Model A transition is marked by a $k^{-2}$ IR divergence for the static correlation function, where $k$ is momentum. This will be put in contrast to the divergences that one gets as a consequence of non-reciprocity. This can similarly be extended to the conserved order parameter Model B variant, which only changes the correlations by an additional factor of $k^2$ \cite{Hohenberg1977, Tauber2014}. The extension to other models in the Hohenberg-Halperin classification we leave for future work. 

Although we consider it on general grounds, such systems occur in a variety of situations. As mentioned before, it can be considered as a set of equations describing the change in concentrations of two reacting and diffusing chemicals, see \cite{Cross1993} and references therein. There, various combinations of relative signs of the constants produce dynamics with the appearance of patterns. Similarly, for models of neural dynamics with excitatory and inhibitory connections, such a model can be used as the lowest order approximation to the dynamics of each sub-population \cite{Cook2022,Wilson1972,Dayan2001, calvo2024}. The constants $m_{1/2}$ then correspond to the natural decay of activity of each sub-population, while the constants $j_{ij}$ correspond to connections between the types, and should have an opposite sign. These are only a few of the examples, with others mentioned in Sec. \ref{sec:intro}.

We now proceed to explore the stability and dynamics of this system.

\subsection{\label{sec:overdamped mean}Mean field}
At the mean field level where we ignore spatial extent and noise in this system, and after defining $\phi_i(x,t) \sim \int d\om e^{i\om t}\hat{\phi}_i(x,\om)$, we see that the dispersion relations for the frequencies are:
\begin{equation}
    i\om_{\pm} = -\f{m_1+m_2}{2} \pm \f{1}{2}\sqrt{(m_1-m_2)^2+4j_{12}j_{21}} \; .
\end{equation}

The system is thus stable when:
\begin{equation}
    m_1m_2 > j_{12}j_{21} = \Delta \; .
\end{equation}

Where we defined the $\Delta$ as a measure of non-reciprocity. That can be seen by introducing $j_{\pm} = (j_{12}\pm j_{21})/2$, see \cite{Fruchart2021}, which allows us to rewrite:
\begin{equation}
    \Delta = j_+^2 - j_-^2 \;.
\end{equation}

We will thus use $\Delta$ as the main control parameter to describe how much non-reciprocity there is in the system. The regions of stability for $\Delta>0$, $\Delta=0$ and $\Delta<0$ are depicted in Fig. \ref{fig:fig1}. Clearly, as non-reciprocity is increased, the region of stability is expanded. We also observe that as we decrease $\Delta$ past zero, instead of a single continuous line marking the edge of stability, we pick up points that separate lines that behave differently. For example, for $\Delta = 0$, where we assume that $j_{21} \neq 0$ without loss of generality, the matrix generating the dynamics at the mean field level is:
\begin{equation}
        \p_t \begin{pmatrix}
            \phi_1 \\ \phi_2
        \end{pmatrix} = - M \begin{pmatrix}
            \phi_1 \\ \phi_2
        \end{pmatrix} = - \begin{pmatrix}
            m_1 & 0 \\ -j_{21} & m_2
        \end{pmatrix}\begin{pmatrix}
            \phi_1 \\ \phi_2
        \end{pmatrix} \,.
\end{equation}

As both $m_1,m_2 \rightarrow 0$, this matrix becomes non-diagonalizable, which marks an Exceptional point (EP). In general, EPs occur when two modes (eigenvalues and eigenvectors) of the system coalesce. Looking at Eq. 2, this can only occur when the modes start to pick up an imaginary part. For $\Delta =0$, this happens at an isolated point, since the modes don't have an imaginary part anywhere within the stable region.  For $\Delta <0$ there are exceptional lines (Fig. 1) that intersect with the critical lines at the corner points. In both cases, points lying on the critical lines that also happen to be exceptional are in fact Critical Exceptional points (CEPs). CEPs occur only outside of equilibrium \cite{Zelle2023}, and exhibit a range of unusual dynamical and statistical properties \cite{Zelle2023, Weis2023, Hanai2020}, some of which we describe next.

\subsection{\label{sec:overdamped statistical}Statistical theory}
The addition of spatial extent in the absence of cross-diffusion simply amounts to modifying $m_i \rightarrow m_i + D_i k^2$, where $k$ is the momentum of the field in Fourier space. Therefore, the stability of each momentum mode can be understood by taking the $(m_1, m_2)$ point, and then moving in a straight line of positive slope from that point, defined by the ratio of $D_1$ and $D_2$ as well as the momentum value $k^2$, see Fig. 1. In the absence of cross-correlations in the noise, the region of stability is not altered. 

In order to calculate the correlation functions and explore the type of transition to instability that each of the critical lines produce, we proceed through the Martin-Siggia-Rose-Janssen-De Dominicis procedure \cite{Altland2023,Kamenev2023,Tauber2014,Martin1973,DeDominicis1978}. Introducing the conjugate fields $\bar{\phi}_1$ and $\bar{\phi}_2$, the action associated with our dynamical system is:
\begin{equation}
    \begin{split}
    &S = \int\, dt d^dx\; i\bar{\phi}_1\left(\p_t \phi_1 +m_1 \phi_1 - j_{12}\phi_2 - D_1 \nabla^2 \phi_1 \right) + \\
    & i\bar{\phi}_2\left(\p_t \phi_2 +m_2 \phi_2 - j_{21}\phi_1 - D_2 \nabla^2 \phi_2  \right)  - \f{B_1}{2}\bar{\phi}_1^2- \f{B_2}{2}\bar{\phi}_2^2 \;.
\end{split}
\end{equation}

From here, the correlation functions can be calculated. As we show in Fig. 1, for most of the lines, the frequency of oscillation between the fields as one approaches critical lines is zero, and it is therefore appropriate to look at the static, or equal time, correlation function in order to understand the transition. We define:
\begin{equation}
    \left<\phi_i(\vec{k}_1, t)\phi_j(\vec{k}_2, t)\right> =\de^d(\vec{k}_1+\vec{k}_2)G_{ij}^K(t=0,\vec{k}_1) \;,
\end{equation}

where the momentum delta function appears due to the spatial translation invariance. However, as shown in Fig. 1, the region between the exceptional lines for the $\Delta <0$ case picks up a non-zero frequency for the oscillation of the field modes. Thus looking at the static correlation function might be inappropriate for the critical line inside that region. We discuss that issue in more detail below. 

\begin{figure*}[!t]
    \includegraphics[width=\textwidth]{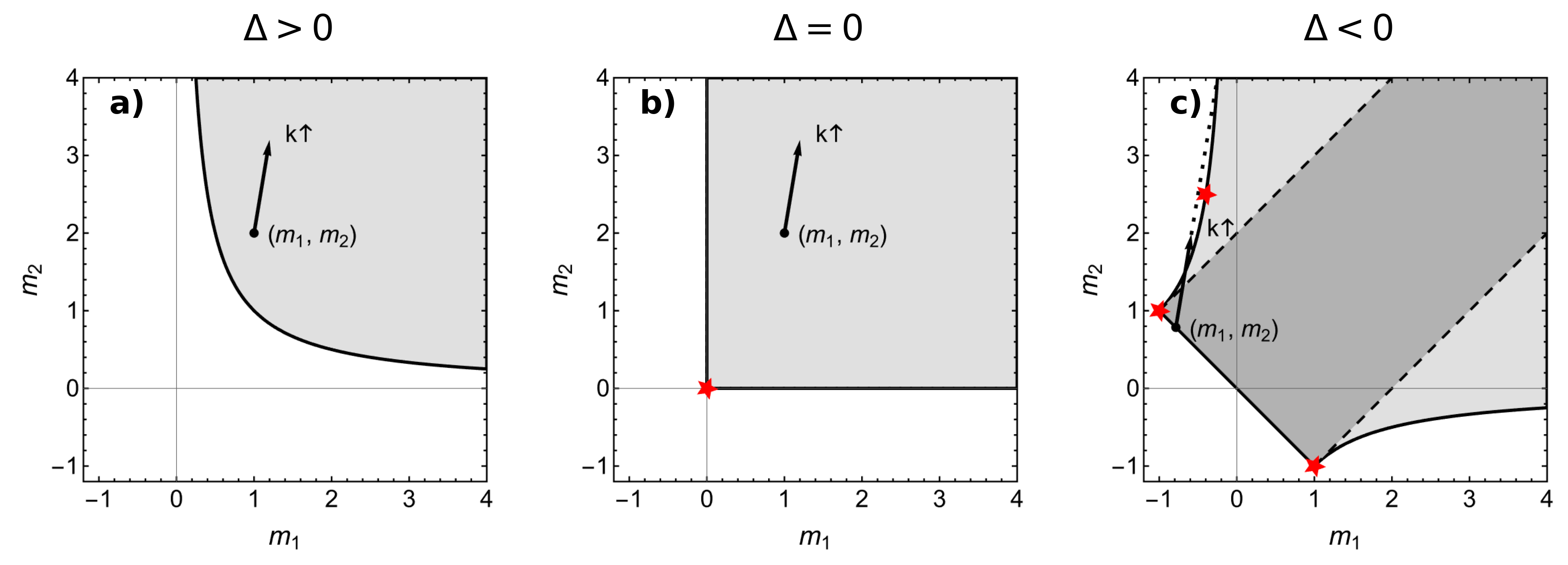}
    \caption{\label{fig:fig1}Stability regions (light grey), oscillating regions (dark grey) and CEPs (stars) of the overdamped system for various $\Delta$. The critical lines are solid, while the dashed lines correspond to the onset of oscillations. The example points $(m_1,m_2)$ with the shift for each mode due to momentum $k$ are plotted as arrows. a) $\Delta>0$. The critical line has the equilibrium $k^{-2}$ divergence in the correlators for small momenta $k$. b) For $\Delta =0$ the critical lines at $m_1=0$ and $m_2=0$ exhibit usual equilibrium behavior, but meet at the CEP when $m_1=m_2=0$. There the correlators pick up stronger $k^{-4}$ and $k^{-6}$ divergences. c) For strongly non-reciprocal regime $\Delta<0$, the region of stability is expanded, and a region with oscillations appears. The critical lines all have $k^{-2}$ divergence in the correlators. However, the line $m_1=-m_2$ is marked by closed rotating orbits in $\phi_1$, $\phi_2$ space. The CEPs where the hyperbola meets the $m_1=-m_2$ lines also exhibit $k^{-4}$ or $k^{-6}$ divergence for equal diffusion constants. The dotted line is chosen with $(m_1,m_2)$ on the critical line with increasing momenta $k$ eventually venturing into the unstable region, thus causing a finite momentum instability. This can be achieved only for high enough $D_2/D_1$. Fig. 2 is plotted along this line.} 
\end{figure*}

The equal-time correlation functions can be calculated exactly for all $\De$:
\begin{widetext}
\begin{subequations}
\begin{align}
    &G_{11}^K(t=0,\vec{k}) \sim \f{j_{12}^2B_2 + B_1\left\{-\De  +\left(m_2 + D_2 k^2\right)\left[m_1+m_2 + \left(D_1+D_2\right)k^2\right]\right\}}{\left[m_1+m_2 + \left(D_1+D_2\right)k^2\right]\left[-\De  +\left(m_2 + D_2 k^2\right)\left(m_1 + D_1 k^2\right)\right]} \;, \\
    &G_{12}^K(t=0,\vec{k}) \sim \f{j_{12}B_2\left(m_1+D_1 k^2\right) + j_{21} B_1\left(m_2 + D_2 k^2\right)}{\left[m_1+m_2 + \left(D_1+D_2\right)k^2\right]\left[-\De  +\left(m_2 + D_2 k^2\right)\left(m_1 + D_1 k^2\right)\right]} \;, \\
    &G_{22}^K(t=0,\vec{k}) \sim \f{j_{21}^2B_1 + B_2\left\{-\De  +\left(m_1 + D_1 k^2\right)\left[m_1+m_2 + \left(D_1+D_2\right)k^2\right]\right\}}{\left[m_1+m_2 + \left(D_1+D_2\right)k^2\right]\left[-\De  +\left(m_2 + D_2 k^2\right)\left(m_1 + D_1 k^2\right)\right]} \;.
\end{align}
\end{subequations}
\end{widetext}

Along the critical lines $m_1m_2 = \De$, the denominator in all three becomes:
\begin{equation}
    k^2\left[m_1+m_2+\left(D_1+D_2\right)k^2\right]\left[D_2m_1 + D_1m_2 + D_1D_2 k^2\right] \;.
\end{equation}

Clearly, for $\De>0$, where both $m_1,m_2>0$ neither of the brackets can be zero in the limit $k\rightarrow 0$, so the correlation functions exhibit the usual divergence of $k^{-2}$ associated with Model A (or $O(2)$ for two fields) dynamics. Interestingly, some non-reciprocity is still allowed to have the equilibrium behavior, as long as the reciprocal coupling is larger.     

When $\De =0$, the critical lines are $m_1=0$ and $m_2=0$. We see that if only one of these is zero, the correlation functions again have the usual $k^{-2}$ divergence. However, if both $m_1=0=m_2$, the correlation functions pick up a $k^{-4}$ and even $k^{-6}$ divergencies (the numerator in $G^{K}_{12}$ reduces the divergence from $k^{-6}$ to $k^{-4}$). For example, if $j_{12}=0$ with $j_{21} \neq 0$ (such that $\De =0$ but there is still non-trivial coupling), it can be seen that $G^{K}_{11}\sim k^{-2}$, $G^{K}_{12}\sim k^{-4}$ and $G^{K}_{22}\sim k^{-6}$ as $k \rightarrow 0$. Such enhanced divergences are a signature of CEPs \cite{Zelle2023, Hanai2020}. Clearly they affect the critical dimensions of the theory from a renormalization group perspective. We see, however, that although such points can be reached without any nonlinearity in the model, it requires tuning of more than one parameters. The phases beyond the critical lines depend on the nonlinearities added to the model, and we leave them for future work.

Finally, for $\Delta < 0$ case, the two critical lines on the hyperbolas exhibit the usual $k^{-2}$ divergence. Here, however, the relative signs of $m_1$ and $m_2$ are different, making it possible that the second bracket in the denominator also contributes to the divergence. This happens when $D_2m_1+D_1m_2=0$. That means that depending on the values of the diffusion constants, there is a CEP along one of the branches of the hyperbola, see Fig. 1. In fact, this point occurs where the line with slope $D_2/D_1$ (which is the line along which we draw modes at different $k$) is tangent to the hyperbola.

Additionally, along the $m_1+m_2=0$ line, we get a different type of divergence. Plugging this into the denominators of the correlation function gives:
\begin{equation}
    k^2 \left(D_1+D_2\right)\left[-\De + \left(-m_1+D_2k^2\right)\left(m_1+D_1k^2\right)\right] \;.
\end{equation}

Therefore this line also exhibits the same divergence as the other critical lines, although it is less intuitive how to interpret it. Usually, the divergence occurs due to the "flattening" of the potential landscape, and some direction becomes undamped. In this case, however, the orbits along this line are closed, see \cite{supplemental}. This is a manifestation of the non-equilibrium nature of the system. The "potential" landscape is flattened in the radial direction, but slanted in the angular direction, hence the closed, rotating orbits. So although it also exhibits a $k^{-2}$ divergence, this critical line is clearly different from the usual equilibrium critical line, since it exhibits rotational dynamics. 

At the intersections of the hyperbola and this line, i.e. where $m_1=-m_2$ and $m_1m_2 = \De$, we get the denominators:
\begin{equation}
    k^4\left(D_1+D_2\right)\left[\left(D_2-D_1\right)m_1 + D_1D_2 k^2\right] \;.
\end{equation}

Thus, these points are CEPs and exhibit the enhanced $k^{-4}$ divergence. We note that the only way to approach this point from the stable region is to have $D_2 \leq D_1$ for $m_1<0$ and $D_2 \geq D_1$ for $m_1>0$, which makes this expression non-negative.  Additionally, if $D_1=D_2$, these points also coincide with the additional CEPs along the hyperbola described above, and the divergence is $k^{-6}$ instead. We can therefore see that one can achieve CEPs by introducing non-reciprocity. Even higher order divergences can be achieved by having more than two fields. The consequences of transitions through these points vary from system to system \cite{Zelle2023,Hanai2020}, but one needs to keep them in mind when nonlinear theories are developed. 

\begin{figure}[t]
\hspace*{-0.05\textwidth}  
\includegraphics[width=0.45\textwidth]{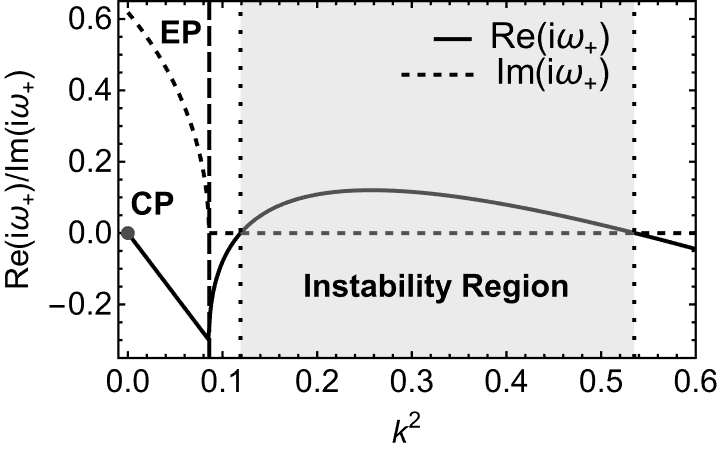} 
\caption{\label{fig:fig2} Re/Im($i\om_+$) vs. $k^2$ for $(m_1,m_2)$ point chosen in Fig. 1, c). $k^2=0$ marks the critical point, where the $Re(i\om_+)=0$, thus marking an instability. In this case the exceptional point occurs where $Im(i\om_+) = 0$, since this is also where $i\om_+ = i\om_-$. As momentum is increased, the modes go into the unstable region, where they would grow indefinitely (unless stabilized by a nonlinearity). This is the finite-momentum instability and is responsible for pattern formation.}
\end{figure}

For $\Delta <0$, we also see that for certain choice of $(m_1,m_2)$ and $D_{1/2}$, the system can develop a finite momentum instability, see Fig. \ref{fig:fig1}. A typical dependence of the least stable mode on momentum $k$ is plotted in Fig. \ref{fig:fig2} for such a choice. Such instabilities are generally stabilized by nonlinear terms, which also determine what kind of patterns the system exhibits. Detailed analyses of pattern formation have been explored in a variety of contexts \cite{Cross1993}, including a nonlinear extension of this model to a non-reciprocal Allen-Cahn model \cite{Liu2023}. Here we pick out several properties.

If the ratio of diffusion constants is such that the line of modes in Fig. 1,c) crosses the critical line at two points, we have two critical momenta $k^{\pm}_c$. The correlation functions at either of those points diverge as $\left|k-k^{\pm}_c\right|^{-1}$. The patterns exhibited by such systems consist of stripes of similar length scales defined by the values of $k$ where instability happens. However, if the ratio of diffusion constants is such that the line of modes is tangential to the critical line for the particular values of $(m_1,m_2)$, the finite momentum instability happens at a CEP (see above). In that case, the divergence of the correlation functions is $\left|k-k_c\right|^{-2}$. We leave the detailed exploration of that transition to future work. 

Finally, one can integrate out one of the fields in order to get the effective dynamics for the other. This can be done by using the dynamical action defined above. We perform this calculation in \cite{supplemental} by integrating out $\phi_2$. Aside from the terms influencing the deterministic behavior of $\phi_1$, we also pick up a term that alters the power spectrum of the noise. In Fourier space the noise correlations are $\left<\xi_1(k_1,\om_1)\xi_1(k_2,\om_2)\right> \sim B_{1}\de(\om_1+\om_2)\de^d(k_1+k_2) $. The effect of integrating out $\phi_2$ amounts to the replacement:
\begin{equation}
    B_1 \rightarrow B_1 + \f{j_{12}^2B_2}{\om_2^2 + (m_2 + D_2 k^2)^2} \;.
\end{equation}

This is similar to the amplification of noise observed in \cite{Biancalani2017}, where it relied on the non-orthogonality of eigenvectors of the matrix generating the dynamics. If we consider the spatially local contribution (which might be the only relevant contribution as the system undergoes some RG flow), we get that in dimensions $d\leq 3$ and for $\om >> m_2$, this additional term becomes:
\begin{equation}
    \int d^dk  \f{j_{12}^2B_2}{\om_2^2 + (m_2 + D_2 k^2)^2} \sim \f{j_{12}^2B_2}{D_2^{\f{d}{2}}\om^{\f{4-d}{2}}} \;.
\end{equation}

We see therefore that $1/f$ noise can be produced simply by coupling the field of interest to another field. The level of non-reciprocity doesn't play a role when the noise is concerned, as long as there is any coupling to the other field. Non-reciprocity will however play a role for the deterministic additions to the dynamics of $\phi_1$, see \cite{supplemental}. However, when $\De=0$, the only contribution is the one in eq. 12. Thus, one can get $1/f$ noise by a one-way coupling to another Gaussian field with spatial correlations.

\section{\label{sec:inertial}Inertial systems}
In the previous section we explored an example overdamped system. That choice was made since often in the study of critical dynamics one approximates the decay of the order parameter towards it's stationary value as relaxational \cite{Tauber2014}. However, for the sake of generality it is prudent to also include the possibility of higher order time derivatives. First of all, real systems often exhibit such dynamics, for example in neural field models \cite{Cook2022}. Additionally, non-equilibrium conditions open up the possibility of the existence of fixed points where the $\p_t^2$ terms are relevant and cannot be discarded \cite{Zelle2023}.

We thus start with a natural extension of the overdamped model:
\begin{equation}
\begin{split}   
    &\p_t^2 \phi_1 + \g_1 \p_t \phi_1 = -m_1 \phi_1 + j_{12} \phi_2 + D_1 \nabla^2 \phi_1 + \xi_1\\
    &\p_t^2 \phi_2 + \g_2 \p_t \phi_2 = -m_2 \phi_2 + j_{21} \phi_1+D_2 \nabla^2 \phi_2 + \xi_2
\end{split}\;.
\end{equation}

Here the terms are defined as in Sec. \ref{sec:overdamped}, with the addition of two real parameters $\g_{1/2}$, which we take to be positive. They can also be taken as negative, and additional terms with cross dependence on $\p_t \phi_i$ can also be introduced. All of these possibilities can change the behavior of the system and should be addressed as needed. However, the number of possible terms that can be included even at the linear level becomes quite large very quickly. Here we restrict ourselves to the above model just to demonstrate that we observe the same effects as in the overdamped system, at least at the static level. 

\subsection{\label{sec:inertial mean}Mean field}
We proceed with the mean field description as before. In Fourier space the frequencies obey the equation:
\begin{equation}
    \begin{split}
    (i\omega)^4 + (\gamma_1 + \gamma_2)(i\omega)^3 + (\gamma_1 \gamma_2 + m_1 + m_2)(i\omega)^2 \\
    + (\gamma_1 m_2 + \gamma_2 m_1)(i\omega) + m_1 m_2 - j_{12} j_{21} = 0 \;.
\end{split}
\end{equation}

Here we again see that the behavior of the system is controlled by the same non-reciprocity parameter $\Delta = j_{12}j_{21}$. Additionally, we see that the relative values of $\g_{1/2}$ should change the dynamics as well, so one can introduce the ratio $ \g_1/\g_2 $ as another control parameter. The effect of diffusion, as before, can be thought of as moving from the point $(m_1,m_2)$ on the stability diagram along a line with positive slope for each mode under consideration. The exact solutions to eq. 15 are bulky, so we will proceed with the $\g_1=\g_2$ case analytically. The other cases can be addressed as needed with the general principles outlined below.

When $\g_1=\g_2=\g$, the least stable solutions are governed by the dispersion:
\begin{equation}
\begin{split}
    &i\om_{\pm} =\\& -\f{\g}{2} + \f{1}{2}\sqrt{\g^2 - 2(m_1+m_2)\pm 2\sqrt{4\Delta + (m_1-m_2)^2}}\; .    
\end{split}
\end{equation}

\begin{figure*}[!t]
    \centering
    \includegraphics[width=\textwidth]{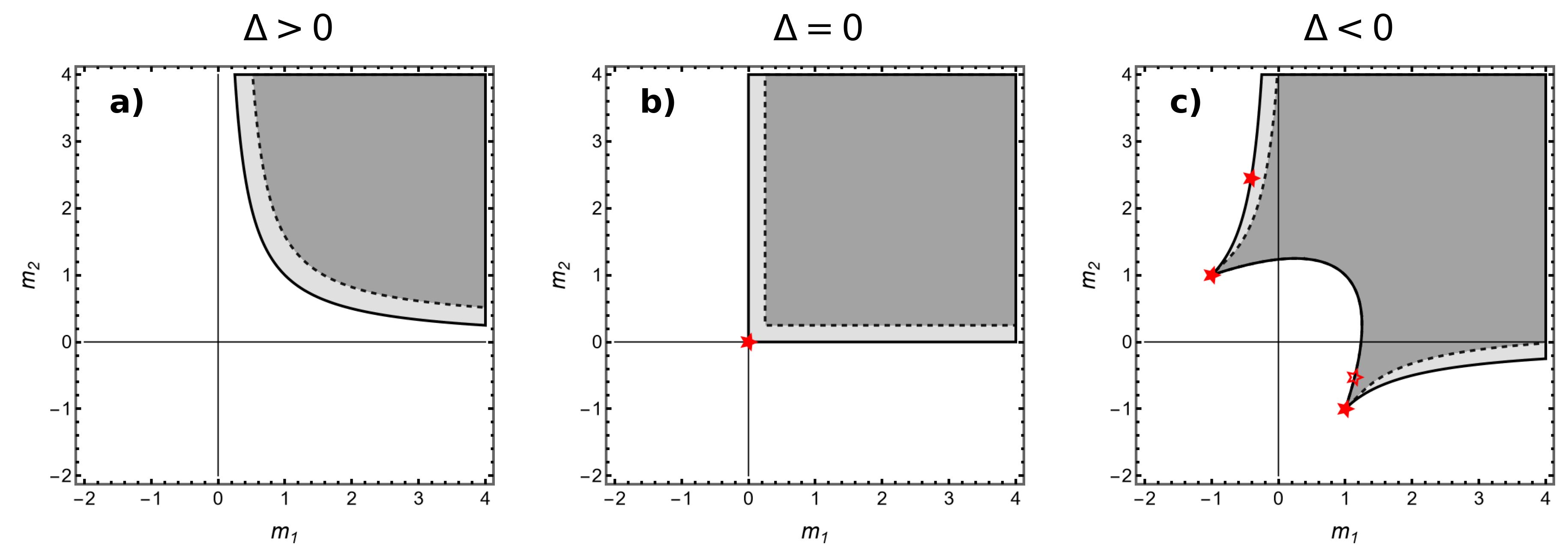}
    \caption{\label{fig:fig3}Stability regions (light grey), regions with oscillations (dark grey) and CEPs (stars) for the inertial system, Eq. 14 for $\g_1=\g_2$. The critical lines are solid, the dashed lines correspond to the onset of oscillations. a) $\Delta>0$. The critical line does not depend on $\g_1/\g_2$, and the correlation functions diverge with decreasing momentum in the usual equilibrium manner as $k^{-2}$. b) $\Delta =0$. The region of stability is the same as in the overdamped case, with the CEP appearing at $m_1=m_2=0$. There the correlators pick up additional $k^{-4}$, $k^{-6}$ divergences. c) $\Delta<0$. Although the region of stability is distorted compared to the overdamped case and now also depends on $\g_1/\g_2$, the qualitative picture is the same. There are now three critical lines, which intersect at two CEPs. The additional CEP along the hyperbola is preserved, and another CEP that appears for sufficiently different $D_1$ and $D_2$ is the hollow star.  There the correlators pick up additional $k^{-4}$ or $k^{-6}$ divergences.} 
\end{figure*}

In Fig. \ref{fig:fig3} we plot the regions of stability as well as the regions for the presence of oscillations for a few example values of $\g_1/\g_2$ and $\Delta$. For $\Delta \geq 0$ the region of stability is the same as for the overdamped case, independent of $\g$. For $\Delta<0$, the region of stability is distorted depending on the values of $\g_1/\g_2$. However, the key features of having three critical lines that create two isolated points still remains. We show below that these points are CEPs. The CEP in the strongly non-reciprocal regime $\De<0$ along the hyperbola remains, and the system also picks up an additional CEP on the internal line. We show all of this in the following section.

\subsection{\label{sec:inertial statistical} Statistical theory}
We can proceed to developing the statistical understanding of this system similarly to the overdamped case. Here we only quote the results for brevity. 

For $\g_1=\g_2=\g$, we again get that the equal time correlation functions are dependent on some denominator. It is given by:
\begin{widetext}
\begin{equation}
        \g \left[\De - \left(m_1+D_1 k^2\right)\left(m_2+D_2 k^2\right)\right]\left[4\De + \left(m_1-m_2+D_1 k^2-D_2k^2\right)^2 +2\g^2\left(m_1+D_1k^2+m_2+D_2k^2\right) \right] \;.
\end{equation}    
\end{widetext}

From the first bracket it is clear that the critical lines are again at $\De = m_1m_2$, which generally diverge as $k^{-2}$. When $\De =0$, we again get $k^{-2}$ divergences on the $m_1 = 0$ and $m_2=0$ lines, and a CEP with $k^{-4}$ and $k^{-6}$ divergences at $m_1=0=m_2$. And also, similarly, when $\De<0$ so $m_1$ and $m_2$ are allowed to take negative values, at the point where $D_2m_1+D_1m_2=0$, we get an enhanced divergence. The second bracket defines an additional line for $\De<0$. The equation for the line is $4\De + (m_1-m_2)^2 + 2\g^2(m_1+m_2)=0$. Along that line, that bracket is:
\begin{equation}
\begin{split}
    k^2& \left[  \left(D_1-D_2\right)^2k^2 \right.\\&\left.+ 2(m_1-m_2)(D_1-D_2)+2\g^2(D_1+D_2)\right]   
\end{split} \;.
\end{equation}

For large enough $D_2/D_1$, the terms not involving $k^2$ inside the brackets can be zero along the this critical line, leading to another CEP. This happens when for some point on the internal critical line the line with slope defined by $D_2/D_1$ is exactly tangent. This is similar to the CEP on the hyperbola. In the same manner, for a choice of $(m_1,m_2)$ close enough to the cusps, and with a high enough ratio of the diffusion constants, another finite momentum instability can occur, now also at a finite frequency.

At the intersections of the two critical lines, we have $\De = m_1m_2$ and $m_1+m_2=0$. In that case the denominators of the correlation functions are:
\begin{equation}
\begin{split}
    &\g k^4 \left[(D_2-D_1)m_1+D_1D_2k^2\right]\\&\left[ \left(D_1-D_2\right)^2k^2 \right.\left.+ 4m_1(D_1-D_2)+2\g^2(D_1+D_2)\right]    
\end{split}\;.
\end{equation}

Thus, we see CEPs with $k^{-4}$ divergences. If the diffusion constants are equal, we see $k^{-6}$ divergences that come from the coalescence of the CEP along the critical line with these points, as in the overdamped case. The case with $\g_1 \neq \g_2$ is not analytically tractable, but the calculations proceed in the same manner, and similar CEPs are present.

Finally, it can be shown similarly to the overdamped case that if one of the fields is eliminated, the other picks up additional noise term. This time, however, the power spectrum of that term goes as $\om^{d-3}$ for dimensions $d\leq 3$. We see again the appearance of $1/f$ noise.

\section{\label{sec:conclusion}Conclusion}
Non-reciprocity is a particular type of deviation from equilibrium behavior. A variety of systems can be thought of as obeying non-reciprocal laws, like spin models, population dynamics, neural behavior and others. Many of these systems can be studied using field-theoretical tools. However, most of these models as well as real phenomena are nonlinear and not solvable exactly. In such cases a field theoretical approach would be to start with a linear description and continue through perturbation series or renormalization group by using that description. It is therefore of practical importance to build sufficient understanding of the linear non-reciprocal models so that they can be used as building blocks for the above-mentioned methods. It turns out that with just two coupled fields, even the linearized dynamics is rich.

We started with a simple reaction-diffusion system with non-reciprocal interactions. However, our results extend to more complex systems, and we included a system with inertial dynamics to show that. We demonstrated that non-reciprocity as we defined it separates different qualitative sets of behaviors. For small non-reciprocity the behavior of the system is essentially equilibrium. It is interesting that some non-reciprocity, and therefore some deviation from equilibrium conditions, is allowed while still retaining equilibrium behavior, at least as far as the statistics of the fields are concerned. It is not clear at this level whether that fact will persist when nonlinearities are added. That would mean that for some small deviations from equilibrium, the long-range behavior still obeys some equilibrium dynamics. In other words, the deviation from equilibrium is irrelevant in the RG sense. We hypothesize that even if possible, such behavior depends strongly on the other elements of the theory. There are, for example, instances where an infinitesimal deviation from equilibrium brings the system to novel fixed points \cite{Daviet2024}.

As the non-reciprocity gets larger, the system acquires isolated points in parameter space where the correlation functions diverge stronger than the usual equilibrium manner. Those are Critical Exceptional Points (CEPs), which are points where two or more modes of the system become equal, while also being critical. The existence of those points is a consequence of the non-equilibrium nature of the model. CEPs are a subject of many ongoing studies, and they display various consequences, from altering the critical dimensions of the model \cite{Hanai2020} to changing the type of transition entirely \cite{Zelle2023,Daviet2024}. We also demonstrate how strong non-reciprocity leads to finite momentum instabilities. In fact, we demonstrate that non-reciprocity needs to be sufficiently strong to achieve that, and that the parameters responsible for the spatial coupling need to be sufficiently different to observe pattern formation. This is in accordance with \cite{Cross1993}, but it means that in most cases CEP's are at least linearly stable to pattern formation.

Finally, we also demonstrated that coupling to a second field produces $1/f$ noise for the dynamics of the first. Various mechanisms from producing such noise are well-studied \cite{milotti20021,Bak1987}, including the mechanism explored in this work. It is possible to produce it even in equilibrium by coupling two fields to each other. 

We believe that in studies of non-reciprocal systems this work should provide the linear, exactly solvable, ground that perturbative calculations can be based on. We leave the exploration of relevant models with the use of results derived here for other works.

\section{Acknowledgements}
This research was partly
supported by the National Science Foundation through
the Center for Living Systems (NSF-PHY 2317138) and by the NSF-MPS-PHY award 2207383.

\nocite{*}
\bibliography{apssamp}

\begin{widetext}
\section*{Supplemental Material}

\subsection{Phase plots}
Here we present the phase plots of the deterministic part of the overdamped system in Eq. 1, without diffusion. For $\Delta >0$, the typical non-critical and critical phase plots are displayed in Fig. \ref{fig:fig1}. When the system is not critical, the fields decay towards the fixed point at $\phi_1=\phi_2=0$. At criticality, some direction becomes "flat", see \ref{fig1:sub2}. Fig. \ref{fig:fig2} shows the phase plots for $\De=0$. The non-critical and general critical behaviors are similar to the $\De>0$ case. However, at the CEP both fields experience a force in some direction, and that direction flips across the fixed point, hence how $\phi_2 =0$ line is also "flat". Finally, Fig. \ref{fig:fig3} shows the typical phase plots for $\De<0$. In \ref{fig3:sub4}, the "flat" direction becomes the radial direction, and the trajectories are closed orbits. 

\begin{figure*}[!h]
    \centering
    \begin{subfigure}{.35\linewidth}
    \centering
    \includegraphics[width=.7\linewidth]{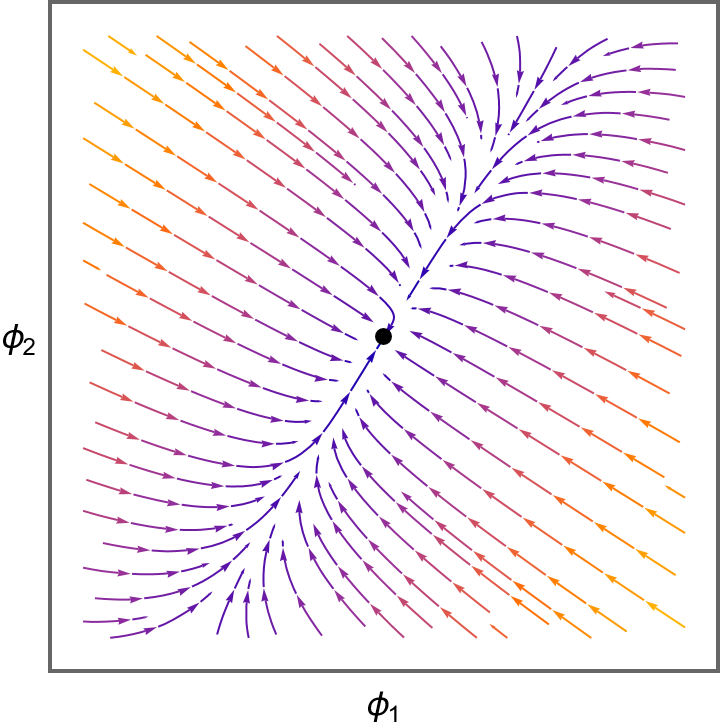}
    \caption{Non-critical}
    \label{fig1:sub1}
    \end{subfigure}%
    \begin{subfigure}{.35\linewidth}
    \centering
    \includegraphics[width=.7\linewidth]{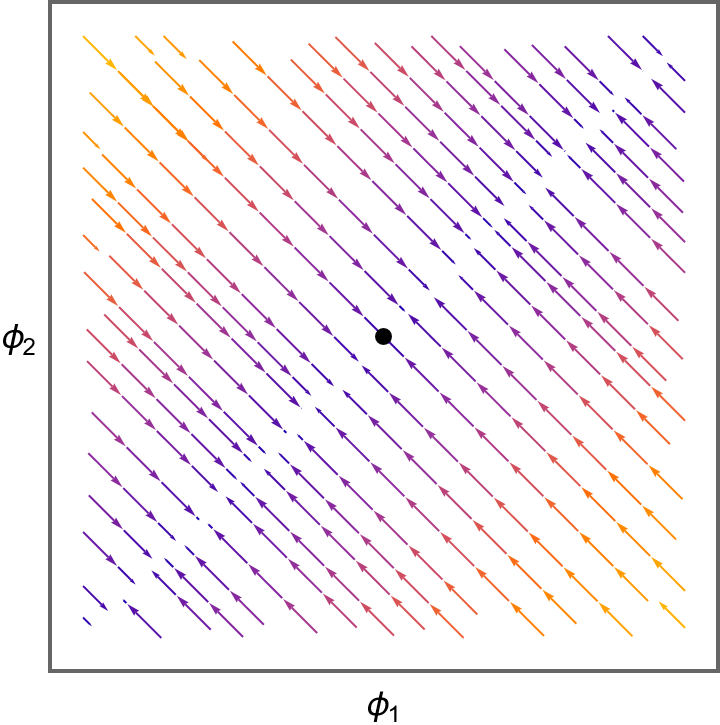}
    \caption{Critical}
    \label{fig1:sub2}
    \end{subfigure}

    \caption{\label{fig:fig4} Typical phase plots for the overdamped system with $\De >0$. The dot corresponds to the fixed point.} 
\end{figure*}

\begin{figure*}[!h]
    \centering
    \begin{subfigure}{.35\linewidth}
    \centering
    \includegraphics[width=.7\linewidth]{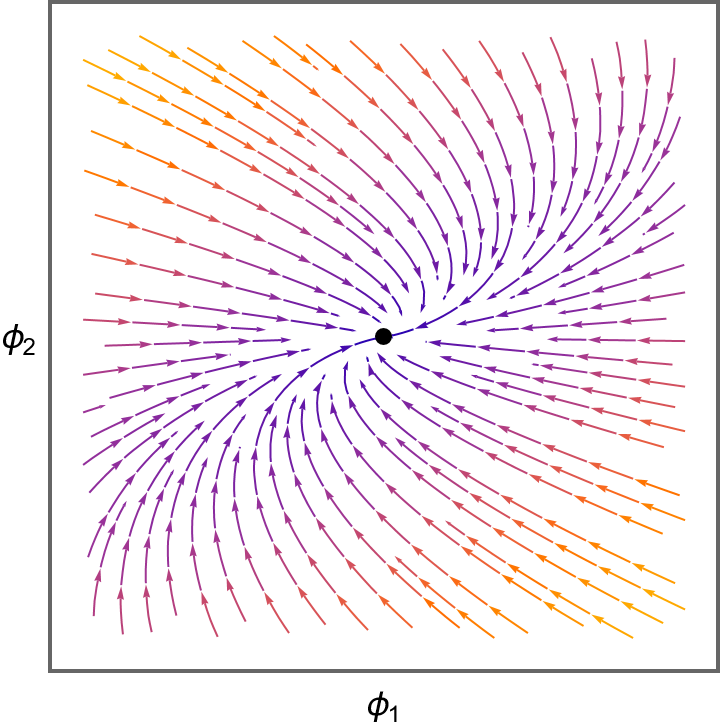}
    \caption{Non-critical}
    \label{fig2:sub1}
    \end{subfigure}%
    \begin{subfigure}{.35\linewidth}
    \centering
    \includegraphics[width=.7\linewidth]{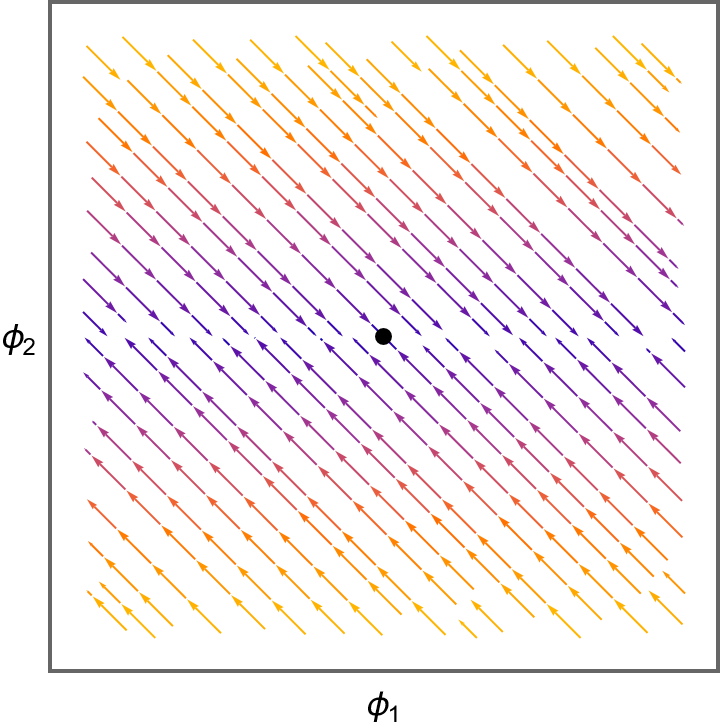}
    \caption{Critical, $m_1 = 0$}
    \label{fig2:sub2}
    \end{subfigure}
    \begin{subfigure}{.35\linewidth}
    \centering
    \includegraphics[width=.7\linewidth]{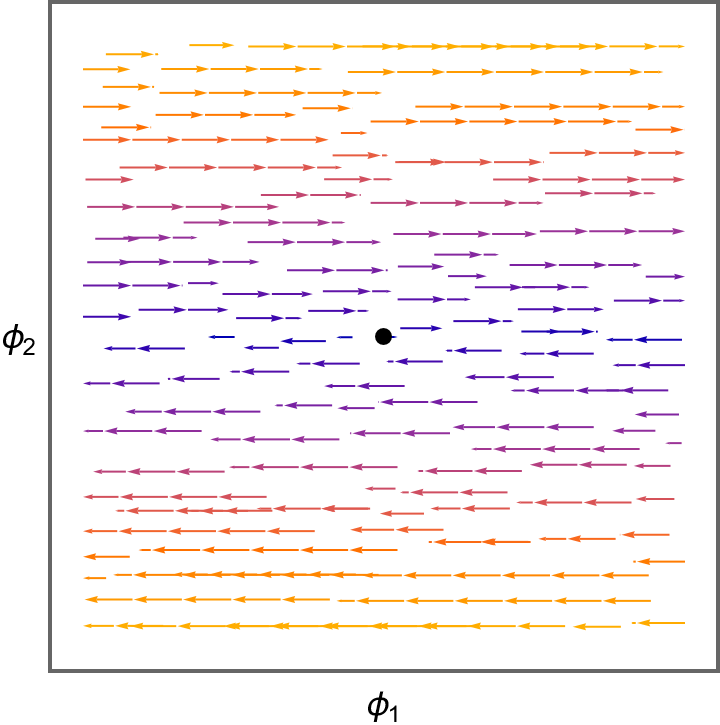}
    \caption{CEP}
    \label{fig2:sub3}
    \end{subfigure}
    \caption{\label{fig:fig5} Typical phase plots for the overdamped system with $\De=0$. The dot corresponds to the fixed point.} 
\end{figure*}

\newpage

\begin{figure*}[!h]
    \centering
    \begin{subfigure}{.4\linewidth}
    \centering
    \includegraphics[width=.8\linewidth]{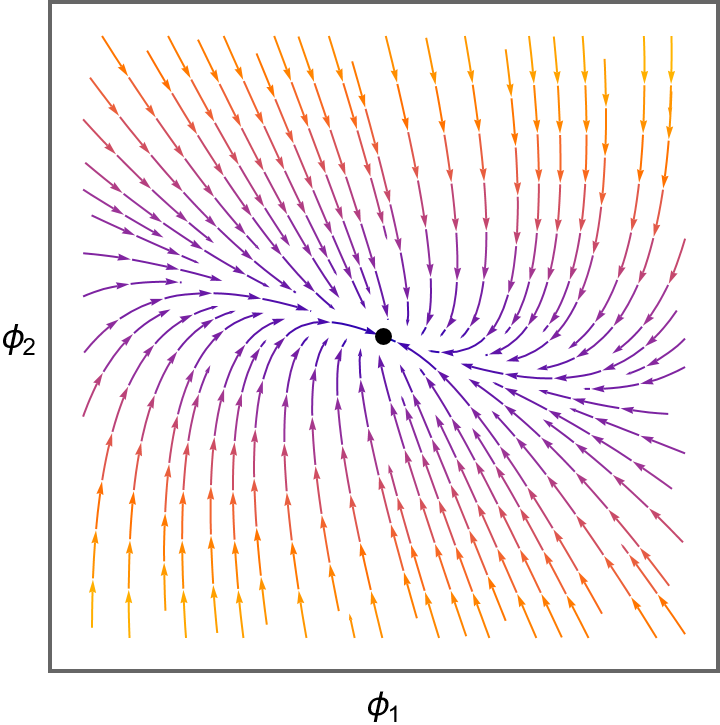}
    \caption{Non-critical, non-rotating}
    \label{fig3:sub1}
    \end{subfigure}%
    \begin{subfigure}{.4\linewidth}
    \centering
    \includegraphics[width=.8\linewidth]{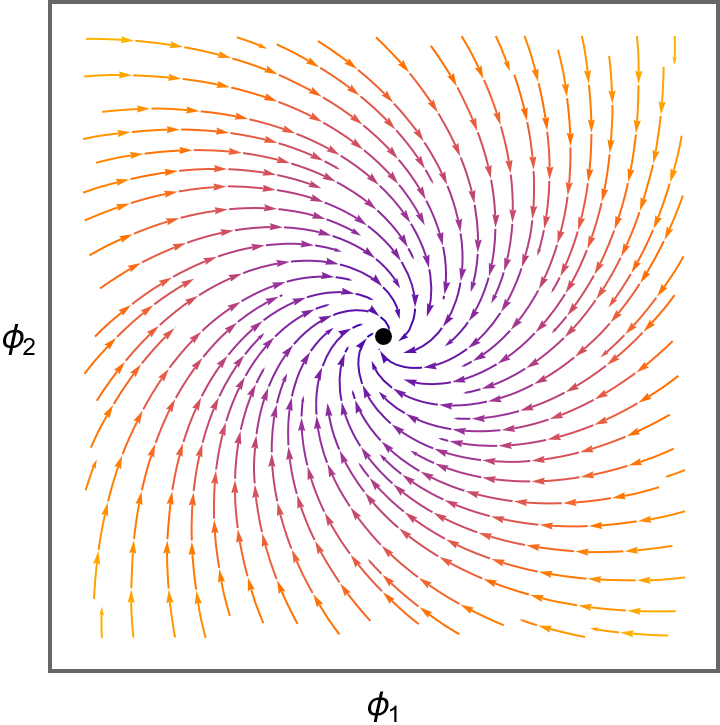}
    \caption{Non-critical, rotating}
    \label{fig3:sub2}
    \end{subfigure}
    
    \begin{subfigure}{.4\linewidth}
    \centering
    \includegraphics[width=.8\linewidth]{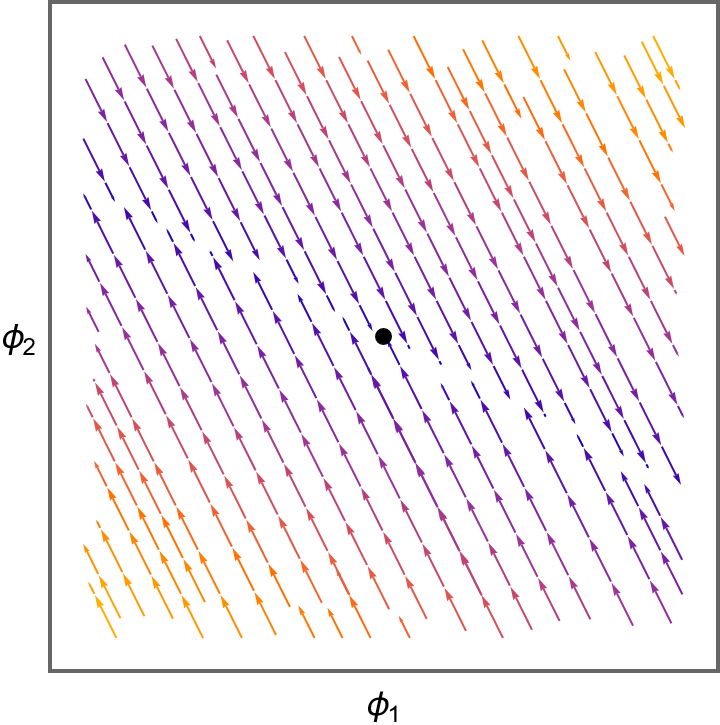}
    \caption{Critical, non-rotating - on the hyperbola}
    \label{fig3:sub3}
    \end{subfigure}%
    \begin{subfigure}{.4\linewidth}
    \centering
    \includegraphics[width=.8\linewidth]{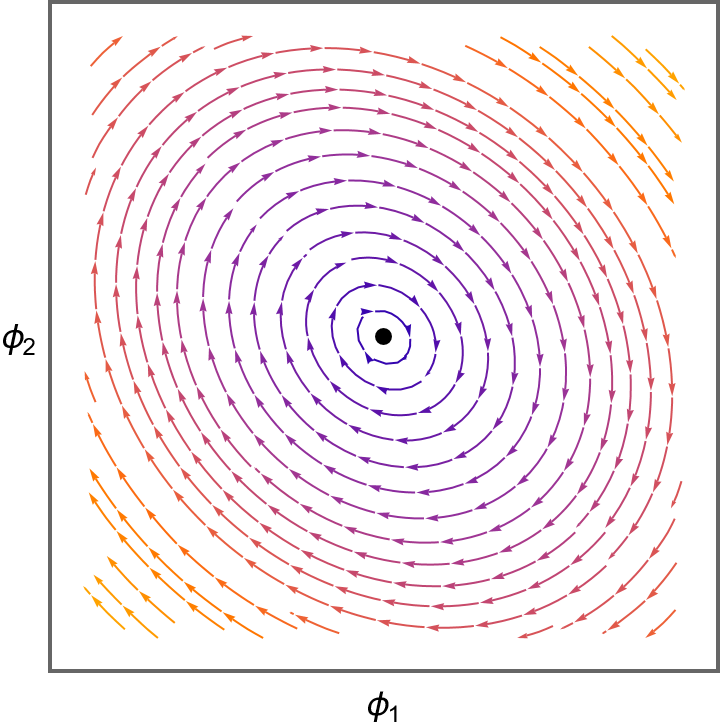}
    \caption{Critical, rotating - on the $m_1=-m_2$ line}
    \label{fig3:sub4}
    \end{subfigure}

    \begin{subfigure}{.4\linewidth}
    \centering
    \includegraphics[width=.8\linewidth]{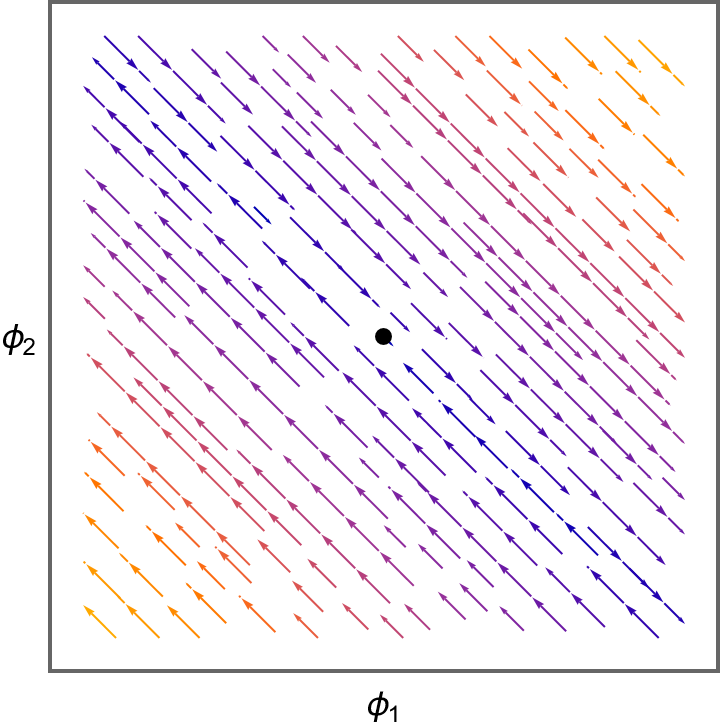}
    \caption{CEP}
    \label{fig3:sub5}
    \end{subfigure}

    \caption{\label{fig:fig6} Typical phase plots for the overdamped system with $\De<0$. The dot corresponds to the fixed point.} 
\end{figure*}

\newpage
\subsection{Effective action of a single field}
Without loss of generality, let's integrate out $\phi_2$. The part of the action that depends on $\phi_2$ can be written as:
\begin{equation}
\begin{split}
    S &= -\f{1}{2}  \int \f{d\om_1d\om_2d^dk_1d^dk_2}{(2\pi)^{2d+2}} (2\pi)^{d+1}\de(\om_1+\om_2)\de^{d}(\vec{k}_1+\vec{k}_2)\boldsymbol{\phi}_2^T(\om_1,\vec{k}_1) G_{0,2}^{-1} \boldsymbol{\phi}_2(\om_2,\vec{k}_2)\\ &+\int \f{d\om d^d k}{(2\pi)^{d+1}} \boldsymbol{J}^T(-\om,-\vec{k})\boldsymbol{\phi}_2(\om,\vec{k})   
\end{split} \; ,
\end{equation}

where $\boldsymbol{\phi}_2(\om,\vec{k}) = \begin{pmatrix}
     \hat{\phi}_2(\om,\vec{k}) & \hat{\bar{\phi}}_2(\om,\vec{k})
\end{pmatrix}^T$ and:
\begin{equation}
    G_{0,2}^{-1}  = \begin{pmatrix}
        0 & -\om_2-im_2-iD_2k_2^2 \\ \om_2-im_2-iD_2k_2^2 & B_2
    \end{pmatrix} \;.
\end{equation} 

Finally: 
\begin{equation}
    \boldsymbol{J}(-\om, -\vec{k}) = \begin{pmatrix}
        -ij_{12}\hat{\bar{\phi}}_1(-\om,-\vec{k}) \\
        -ij_{21}\hat{\phi}_1(-\om,-\vec{k})
    \end{pmatrix} \;.
\end{equation}

Integrating it out then yields:
\begin{equation}
\begin{split}
    &\f{1}{2}\int \f{d\om_1d\om_2 d^dk_1d^dk_2}{(2\pi)^{2d+2}}(2\pi)^{d+1}\de(\om_1+\om_2)\de^{d}(\vec{k}_1+\vec{k}_2) J(\om_1,\vec{k}_1)G_{0,2}J(\om_2,\vec{k}_2) \\&=\f{1}{2}\int \f{d\om_1d\om_2 d^dk_1d^dk_2}{(2\pi)^{2d+2}}(2\pi)^{d+1}\de(\om_1+\om_2)\de^{d}(\vec{k}_1+\vec{k}_2) \left[-j_{12}^2\hat{\bar{\phi}}_1(\om_1,\vec{k}_1)\f{B_2}{\om_2^2+(m_2+D_2k_2^2)^2}\hat{\bar{\phi}}_1(\om_2,\vec{k}_2) \right. \\&-\left. j_{12}j_{21}\hat{\bar{\phi}}_1(\om_1,\vec{k}_1) \f{1}{\om_2 - im_2-iD_2k_2^2}\hat{\phi}_1(\om_2,\vec{k}_2) - j_{12}j_{21}\hat{\bar{\phi}}_1(\om_2,\vec{k}_2) \f{1}{-\om_2 - im_2-iD_2k_2^2}\hat{\phi}_1(\om_1,\vec{k}_1)\right]
\end{split}
\end{equation}

The deterministic terms are determined by $\De = j_{12}j_{21}$ and depend on the response functions. The noise term is equivalent to adding to the noise kernel of $\phi_1$ an additional term:
\begin{equation}
    \f{j_{12}^2 B_2}{\om^2 + (m_2 + D_2 k^2)^2}
\end{equation}

\end{widetext}

\end{document}